\documentclass[12pt]{article}

\input epsf

\makeatletter
\def\numberbysection{\@addtoreset{equation}{section}
 	\def\theequation{\thesection.\arabic{equation}}}
\makeatother

\numberbysection

\newcommand{\be}{\begin{eqnarray}}
\newcommand{\ee}{\end{eqnarray}}
\newcommand{\non}{\nonumber}

\newcommand{\n}{\ensuremath{\mathcal{N}}}

\begin{document}

\begin{titlepage}
\strut\hfill UMTG--218
\vspace{.5in}
\begin{center}

\LARGE Direct Calculation of Breather $S$ Matrices \\[1.0in]
\large Anastasia Doikou and Rafael I. Nepomechie\\[0.8in]
\large Physics Department, P.O. Box 248046, University of Miami\\[0.2in]  
\large Coral Gables, FL 33124 USA\\

\end{center}

\vspace{.5in}

\begin{abstract}
    We formulate a systematic Bethe-Ansatz approach for computing 
    bound-state (``breather'') $S$ matrices for integrable quantum spin 
    chains.  We use this approach to calculate the breather boundary $S$ 
    matrix for the open XXZ spin chain with diagonal boundary fields.  We 
    also compute the soliton boundary $S$ matrix in the critical regime.
\end{abstract}

\end{titlepage}

\section{Introduction}

A common feature of integrable models is the existence of bound states 
for a certain range of the coupling constant.  A well-known example is 
the sine-Gordon/massive Thirring model, which in the attractive 
regime ($0 < \beta^{2} < 4 \pi$) exhibits soliton-antisoliton bound 
states called ``breathers.''  (See e.g.  \cite{ZZ} and references 
therein.) The direct Bethe-Ansatz calculation of exact scattering 
matrices for both solitons (also known as ``kinks'' or ``holes'') and 
breathers was pioneered by Korepin \cite{korepin}.  Andrei and Destri 
\cite{AD} later systematized such $S$ matrix calculations for the 
solitons.  We develop here a corresponding systematic approach for 
computing $S$ matrices for the breathers.  In particular, we give a 
direct calculation of the breather boundary $S$ matrix for the open 
XXZ spin chain with diagonal boundary fields \cite{ABBBQ},\cite{sklyanin}.  
Our results coincide with the bootstrap results for the boundary 
sine-Gordon model \cite{GZ} with ``fixed'' boundary conditions which 
were obtained by Ghoshal \cite{ghoshal}.  We also give a direct 
computation of the soliton boundary $S$ matrix in the critical regime 
\cite{GZ},\cite{FS} using the method developed in 
\cite{GMN},\cite{DMN}.  Although we focus on the XXZ chain, we expect 
that our method of computing bound-state $S$ matrices should be 
applicable to other integrable quantum spin chains.

Bulk calculations are generally more straightforward than 
corresponding boundary calculations.  We therefore first formulate in 
Section 2 the method of computing breather $S$ matrices for the case 
of bulk (two-particle) scattering in the closed XXZ chain, and thereby 
reproduce the well-known results \cite{ZZ}, \cite{korepin}, \cite{KR}.  
In Section 3 we turn to the open XXZ chain.  We compute the breather 
boundary $S$ matrix, and find agreement with the bootstrap results 
provided a certain identification of boundary parameters is made.  In
order to further check this identification, we also compute the soliton 
boundary $S$ matrix. A brief comparison of our approach with that of 
other authors is given in Section 4.

\section{Closed XXZ chain}

In this Section we consider the periodic anisotropic Heisenberg 
(or ``closed XXZ'' ) spin chain in the critical regime, whose 
Hamiltonian is given by \cite{KR},\cite{dCG},\cite{TS}
\be
{\cal H} =  {\epsilon\over 4} \sum_{n=1}^{N} \left\{
\sigma^{x}_n \sigma^{x}_{n+1}
+ \sigma^{y}_n \sigma^{y}_{n+1} 
+ \cos \mu  \left( \sigma^{z}_n \sigma^{z}_{n+1} - 1 \right) \right\} \,, 
\qquad \vec \sigma_{N+1} \equiv \vec \sigma_{1} \,, 
\label{hamiltonian} 
\ee 
with $0 < \mu < {\pi\over 2}$ and $\epsilon=\pm 1$. We also assume 
that the number of spins $(N)$ is even. It can be 
shown (see e.g.  \cite{FS},\cite{KR},\cite{DN1}) that the 
kink $S$ matrix coincides with the sine-Gordon soliton $S$ matrix 
\cite{ZZ}, provided that the sine-Gordon coupling constant $\beta^{2}$ 
is identified as
\be
\beta^{2} = \left\{  \begin{array}{cl}
8 \left( \pi - \mu \right) & \mbox{for} \quad  \epsilon= +1  \\
8 \mu                      & \mbox{for} \quad  \epsilon= -1
\end{array} \right.
\,.
\label{identifications}
\ee 
Since we restrict the anisotropy parameter $\mu$ to the range $(0 \,, 
{\pi\over 2})$, it follows that the case $\epsilon= +1$ corresponds to 
the ``repulsive'' regime ($4\pi < \beta^{2} < 8 \pi$) of the 
sine-Gordon model in which there are no bound states, while $\epsilon= 
-1$ corresponds to the ``attractive'' regime ($0 < \beta^{2} < 4 \pi$) 
in which bound states do exist.

Choosing the pseudovacuum to be the ferromagnetic state with all spins 
up, the algebraic Bethe Ansatz \cite{ABA} can be used to construct
simultaneous eigenstates of the Hamiltonian, momentum, and $S^{z}$. The 
corresponding eigenvalues are given by \footnote{The dependence on
$\epsilon$ in the formulas that follow is explained in Appendix A.}
\be
E &=& - \epsilon \sin^{2} \mu  \sum_{\alpha=1}^{M} 
{1\over \cosh (2 \mu \lambda_{\alpha}) - \cos \mu } \,, \label{energy} \\ 
P &=& \pi M \theta(-\epsilon) + {\epsilon \over i} \sum_{\alpha=1}^{M} 
\log {\sinh  \mu \left( \lambda_{\alpha} + {i\over 2} \right) 
\over \sinh  \mu \left( \lambda_{\alpha} - {i\over 2} \right)} 
\quad (\mbox{mod } 2 \pi) \,, \label{momentum} \\ 
S^{z }&=& {N\over 2} - M \,, \label{spin}
\ee 
where $\{ \lambda_{1} \,, \ldots \,, \lambda_{M} \}$ are solutions of
the Bethe Ansatz equations
\be 
e_{1}(\lambda_{\alpha}\,; \mu)^{N} = 
\prod_{\scriptstyle{\beta=1}\atop \scriptstyle{\beta \ne \alpha}}^M 
e_{2}(\lambda_{\alpha}-\lambda_{\beta}\,; \mu) \,, \quad \alpha = 1 
\,, \cdots \,, M \,,
\label{BAE}
\ee
where
\be
e_{n}(\lambda\,; \mu) = 
{\sinh \mu \left( \lambda + {i n\over 2} \right) 
\over \sinh \mu \left( \lambda - {i n\over 2} \right) } \,.
\ee
Moreover, $\theta(x)$ is the Heaviside unit step function.

For the analysis that follows, it is convenient to also introduce the following 
notations:
\be
g_{n}(\lambda \,; \mu) = e_{n}(\lambda \pm {i \pi \over 2 \mu}\,; \mu)
= {\cosh \mu \left( \lambda + {i n\over 2} \right) 
\over \cosh \mu \left( \lambda - {i n\over 2} \right) } \,,
\ee 
\be
q_n (\lambda\,; \mu) &=& \pi + i\log e_n(\lambda\,; \mu) \,, \non \\
r_n (\lambda\,; \mu) &=& i\log g_n(\lambda\,; \mu) \,,
\label{qr}
\ee
\be
a_n(\lambda\,; \mu) &=& {1\over 2\pi} {d \over d\lambda} q_n (\lambda\,; \mu)
= {\mu \over \pi} 
{\sin (n \mu)\over \cosh(2 \mu \lambda) - \cos (n \mu)} \,, \non \\
b_n(\lambda\,; \mu) &=& {1\over 2\pi} {d \over d\lambda} r_n (\lambda\,; \mu)
= -{\mu \over \pi} 
{\sin (n \mu)\over \cosh(2 \mu \lambda) + \cos (n \mu)} \,. 
\ee 
The latter functions, which have the periodicity $n \rightarrow n + 
{2\pi\over \mu}$, have the following Fourier transforms \footnote{Our 
conventions are
\be
\hat f(\omega) \equiv \int_{-\infty}^\infty e^{i \omega \lambda}\ 
f(\lambda)\ d\lambda \,, \qquad\qquad
f(\lambda) = {1\over 2\pi} \int_{-\infty}^\infty e^{-i \omega \lambda}\ 
\hat f(\omega)\ d\omega \,, \non 
\ee 
and we use $*$ to denote the convolution
\be
\left( f * g \right) (\lambda) = \int_{-\infty}^\infty 
f(\lambda - \lambda')\ g(\lambda')\ d\lambda' \,. \non 
\ee  
}:
\be
\hat a_{n}(\omega\,; \mu) = {\sinh \left( (\nu  - n) 
{\omega \over 2} \right) \over
\sinh \left( {\nu \omega \over 2} \right)} \,,
\qquad 0 < n < 2 \nu  \,, 
\label{fourier1}
\ee
\be
\hat b_{n}(\omega\,; \mu) &=&
-{\sinh \left( {n \omega \over 2} \right) \over
\sinh \left( {\nu \omega \over 2} \right)} \,,
\qquad \qquad \quad  0 < n < \nu  \,, \label{fourier2} \\
&=& -{\sinh \left( ( n - 2 \nu) {\omega \over 2} \right) \over
\sinh \left( {\nu \omega \over 2} \right)} \,,
\qquad  \nu  < n < 2 \nu \,, \label{fourier3}
\ee
where $\nu = {\pi \over \mu} > 2$.

\subsection{Ground state}

In order to study the breathers, we must consider the attractive case 
$\epsilon= -1$.  The ground state lies in the sector with $N$ even, 
and is characterized by a ``sea'' of $M={N\over 2}$ negative-parity 
1-strings (i.e., roots of the form $\lambda^{0} + {i \pi \over 2 
\mu}$, where the ``center'' $\lambda^{0}$ is real) \cite{TS}.  We 
briefly review the procedure for determining the root density, which 
describes the distribution of roots in the thermodynamic ($N 
\rightarrow \infty$) limit.  The Bethe Ansatz Eqs.  (\ref{BAE}) for 
the ground state are
\be 
g_{1}(\lambda_{\alpha}\,; \mu)^{N} = 
\prod_{\scriptstyle{\beta=1}\atop \scriptstyle{\beta \ne \alpha}}^M 
e_{2}(\lambda_{\alpha}-\lambda_{\beta}\,; \mu) \,, \quad \alpha = 1 
\,, \cdots \,, M \,,
\ee
with $\{ \lambda_{\alpha} \}$ {\it real}. By taking logarithms, these 
equations can be rewritten as
\be 
h (\lambda_\alpha) = J_\alpha \,,  
\qquad \alpha = 1 \,, \cdots \,, M \,,
\ee 
where the so-called counting function $h(\lambda)$ is given by
\be 
h(\lambda) = -{1\over 2\pi} \left\{  N r_1(\lambda\,; \mu) 
- \sum_{\beta=1}^{M}  q_{2}(\lambda - \lambda_{\beta}\,; \mu) 
\right\} \,, 
\label{counting} 
\ee 
and $\{ J_\alpha \}$ are certain integers or half-integers.  The sign 
of the counting function is chosen so as to make it a monotonically
increasing function of $\lambda$.
The root density $\sigma(\lambda)$ is defined by
\be
\sigma(\lambda) = {1\over N} {d \over d \lambda} h(\lambda) \,,
\label{sigma}
\ee 
so that the number of $\lambda_{\alpha}$ in the interval $[ \lambda 
\,, \lambda + d\lambda ]$ is $N \sigma(\lambda) d\lambda$. It is a 
positive function by virtue of the monotonicity of the counting 
function. Passing from the sum in $h(\lambda)$ to an integral, we obtain a
linear integral equation for the root density
\be
\sigma(\lambda) = -b_{1}(\lambda \,; \mu) 
+ \int_{-\infty}^{\infty} d\lambda' \sigma(\lambda')\ 
a_{2}(\lambda - \lambda'\,; \mu) \,.
\ee 
Solving this equation by Fourier transforms using 
Eqs. (\ref{fourier1}),(\ref{fourier2}) , we 
conclude that the root density for the ground state is given by
\be
\sigma(\lambda) = s(\lambda) = {1\over 2\pi} 
\int_{-\infty}^\infty d\omega\ e^{-i \omega \lambda}\ \hat s(\omega)
 = {1\over 2 (\nu - 1) \cosh \left( {\pi \lambda \over \nu - 1} \right)} \,,
\label{density/ground}
\ee 
where
\be
\hat s(\omega) = {\hat b_{1}(\omega\,; \mu)
\over -1 + \hat a_{2}(\omega\,; \mu)} = 
{1\over 2 \cosh \left( (\nu - 1) {\omega\over 2} \right)}
\,.
\ee 

We verify the consistency of this procedure by computing the value of 
$M$ from the root density:
\be
M = \sum_{\alpha=1}^{M} 1 
= N \int_{-\infty}^{\infty}d\lambda\ \sigma(\lambda) 
= N \hat s (0) = {N\over 2} \,,
\label{consistency}
\ee
and hence, the state indeed has $S^{z}=0$. The energy and momentum are
\be
E_{gr} &=& {\pi \sin \mu \over \mu } \sum_{\alpha=1}^{M}
a_{1}(\lambda_{\alpha} + {i \pi \over 2 \mu}\,; \mu) 
= {\pi \sin \mu \over \mu } \sum_{\alpha=1}^{M} 
b_{1}(\lambda_{\alpha}\,; \mu) 
= {\pi \sin \mu \over \mu }
N \int_{-\infty}^{\infty} d\lambda\ s(\lambda)\ b_{1}(\lambda \,; \mu) \non \\
&=& -{\sin \mu \over 4 \mu } N \int_{-\infty}^{\infty} d\omega
{\sinh \left( {\omega \over 2} \right) \over
\cosh \left( (\nu - 1) {\omega\over 2} \right)
\sinh \left( {\nu \omega \over 2} \right)} \,, \non \\
P_{gr} &=& \pi M  + \sum_{\alpha=1}^{M} \left[ 
q_{1} ( \lambda_{\alpha}+ {i \pi \over 2 \mu}\,; \mu )  - \pi \right] 
= \pi M + \sum_{\alpha=1}^{M} r_{1}(\lambda_{\alpha}\,; \mu) 
= \pi M + N \int_{-\infty}^{\infty} d\lambda\ s(\lambda)\ 
r_{1}(\lambda\,; \mu) \non \\
&=& {\pi N\over 2} \quad (\mbox{mod } 2 \pi) \,.  
\ee 

\subsection{Two-breather state}

As for the massive Thirring/sine-Gordon model \cite{korepin}, the XXZ 
chain in the attractive regime has two classes of excitations above 
the ground-state sea: holes which correspond to solitons, and strings 
which correspond to soliton-antisoliton bound states, i.e., breathers.  
The $n^{th}$ breather corresponds to a positive-parity $n$-string; 
i.e., a set of $n$ roots of the Bethe Ansatz Eqs.  of the form
\be
\lambda^{(n ,l)} = \lambda^{0} + {i\over 2}\left( n + 1 - 2l \right) \,,
\qquad l = 1 \,, \ldots \,, n \,,
\ee
where the center $\lambda^{0}$ is real.  In particular, the 
fundamental breather ($n=1$) corresponds to a real root of the Bethe 
Ansatz equations.  Breather states exist only for $n \in \{ 1 \,, 
\ldots \,, [\nu] - 1 \}$, where $[x]$ denotes integer part of $x$ (see 
\cite{korepin},\cite{TS}).

We consider now an excited state consisting of two breathers 
$\lambda_{1}^{(n_{1},l_{1})}$, $\lambda_{2}^{(n_{2},l_{2})}$ (with 
centers $\lambda^{0}_{1}$ and $\lambda^{0}_{2}$, respectively) in the 
sea, again with $N$ even.  The Bethe Ansatz Eqs. (\ref{BAE}) now imply
\be 
g_{1}(\lambda_{\alpha}\,; \mu)^{N} &=&
-\prod_{\beta=1}^{M^{-}_{1}} 
e_{2}(\lambda_{\alpha}-\lambda_{\beta}\,; \mu) 
\prod_{\beta=1}^{2} \prod_{l_{\beta}=1}^{n_{\beta}}
g_{2}(\lambda_{\alpha}- \lambda_{\beta}^{(n_{\beta},l_{\beta})}\,; \mu)
\,, \quad \alpha = 1 \,, \cdots \,, M^{-}_{1} \,, \non  \\ 
\label{bulk/sea} \\
e_{1}(\lambda_{1}^{(n_{1},l_{1})}\,; \mu)^{N} &=&
\prod_{\beta=1}^{M^{-}_{1}} 
g_{2}(\lambda_{1}^{(n_{1},l_{1})} -\lambda_{\beta}\,; \mu)
\prod_{l_{2}=1}^{n_{2}}
e_{2}(\lambda_{1}^{(n_{1},l_{1})} - \lambda_{2}^{(n_{2},l_{2})} \,; \mu)
\,, \quad l_{1} = 1 \,, \cdots \,, n_{1} \,, \non  \\ 
\label{bulk/string}
\ee
where $M^{-}_{1}$ is the number of roots in the sea, and $\lambda_{1} 
\,, \ldots \,, \lambda_{M^{-}_{1}}$ are real.

The first set of equations (\ref{bulk/sea}), which describes the 
(distorted) sea, implies the counting function
\be
h(\lambda) = -{1\over 2\pi} \left\{  N r_1(\lambda\,; \mu) 
- \sum_{\beta=1}^{M^{-}_{1}}  q_{2}(\lambda - \lambda_{\beta}\,; \mu) 
- \sum_{\beta=1}^{2} \sum_{l_{\beta}=1}^{n_{\beta}}
r_{2}(\lambda - \lambda_{\beta}^{(n_{\beta},l_{\beta})}\,; \mu)
\right\} \,. 
\ee 
The corresponding root density (\ref{sigma}) is therefore given by
\be
\sigma(\lambda) &=& s(\lambda) 
- {1\over N} \sum_{\beta=1}^{2} \sum_{l_{\beta}=1}^{n_{\beta}}
K_{1}(\lambda - \lambda_{\beta}^{(n_{\beta},l_{\beta})})  
\label{sigma1} \\
&=& s(\lambda) - {1\over N}\sum_{\beta=1}^{2} 
K_{n_{\beta}}(\lambda - \lambda_{\beta}^{0}) \,, \label{sigma2}
\ee 
where the Fourier transform of $K_{n}(\lambda)$ is given by
\be
\hat K_{n}(\omega) = 
{\hat b_{n-1}(\omega \,; \mu) + \hat b_{n+1}(\omega \,; \mu) \over
-1 + \hat a_{2}(\omega \,; \mu)} =
{\sinh \left( {n \omega \over 2} \right) 
\cosh  \left( {\omega \over 2} \right) \over 
\sinh \left( {\omega \over 2} \right) 
\cosh  \left( (\nu -1) {\omega \over 2} \right)} \,,
\label{Kn}
\ee
keeping in mind that $n < \nu - 1$.
A calculation analogous to (\ref{consistency}) shows that
$M^{-}_{1}={N\over 2} - n_{1} - n_{2}$, and therefore, the breathers 
have $S^{z}=0$. The energy of the state is given by
\be
E &=& {\pi \sin \mu \over \mu } \left\{ \sum_{\alpha=1}^{M^{-}_{1}}
a_{1}(\lambda_{\alpha} + {i \pi \over 2 \mu}\,; \mu) 
+ \sum_{\alpha=1}^{2} \sum_{l_{\alpha}=1}^{n_{\alpha}} 
a_{1}( \lambda_{\alpha}^{(n_{\alpha},l_{\alpha})}\,; \mu) \right\}
\non \\ 
&=& {\pi \sin \mu \over \mu } \left\{
N \int_{-\infty}^{\infty} d\lambda\ \sigma (\lambda)\ b_{1}(\lambda \,; \mu) 
+ \sum_{\alpha=1}^{2} a_{n_{\alpha}}( \lambda_{\alpha}^{0}\,; \mu) \right\}
\non \\
&=& E_{gr} + {\pi \sin \mu \over \mu } \sum_{\alpha=1}^{2}
\varepsilon_{n_{\alpha}}( \lambda_{\alpha}^{0}) \,, 
\ee 
where the Fourier transform of $\varepsilon_{n}(\lambda)$ is given by
\be
\hat \varepsilon_{n}(\omega) = \hat a_{n}(\omega \,; \mu) 
- \hat K_{n}(\omega) \hat b_{1}(\omega \,; \mu) =
{\cosh  \left( (\nu - n - 1) {\omega \over 2} \right) \over 
\cosh  \left( (\nu -1) {\omega \over 2} \right)} \,,
\ee 
which is invariant under $n \rightarrow -n + 2 (\nu -1)$.
Similarly, the momentum of the state is given by
\be
P &=& \pi M^{-}_{1}
+ N \int_{-\infty}^{\infty} d\lambda\ \sigma (\lambda)\ r_{1}(\lambda \,; \mu) 
+ \sum_{\alpha=1}^{2} q_{n_{\alpha}}( \lambda_{\alpha}^{0}\,; \mu) \non \\
&=& P_{gr} + \sum_{\alpha=1}^{2} p_{n_{\alpha}}( \lambda_{\alpha}^{0}) \,, 
\ee 
where the breather momentum $p_{n}(\lambda)$ is given by
\be
p_{n}(\lambda) = -\left( K_{n} * r_{1} \right) (\lambda) + 
q_{n}(\lambda \,; \mu) \,,
\label{breathermomentum}
\ee
It is now easy to verify the important relation
\be
{1\over 2 \pi} {d\over d\lambda} p_{n}(\lambda) = \varepsilon_{n}(\lambda)
\,.
\label{energy/momentum}
\ee 

We remark that the following bootstrap-like relations are easily 
verified \cite{KR}:
\be
\varepsilon_{j}(\lambda + {i\over 2}k ) + 
\varepsilon_{k}(\lambda - {i\over 2}j ) &=& \varepsilon_{j+k}(\lambda) 
\,, \non \\
s(\lambda + {i\over 2}(\nu - 1 - j)) + 
s(\lambda -{i\over 2}(\nu - 1 - j)) &=& \varepsilon_{j}(\lambda) \,.
\ee 
Indeed, a hole (soliton) with rapidity $\lambda$ can be shown 
to have energy ${\pi \sin \mu \over \mu }  s(\lambda)$. We also 
remark that charge conjugation (${\cal C}$) and parity (${\cal P}$) 
eigenvalues can be readily computed using the methods described in 
Ref. \cite{DN1}. Indeed, we find that the ground state is an 
eigenstate of ${\cal C}$ and ${\cal P}$ with eigenvalue $(-)^{N\over 2}$.
Moreover, an $n$-breather state has ${\cal C}=(-)^{{N\over 2}-n}$; and 
if the rapidity is zero this state is also a parity eigenstate with 
${\cal P}=(-)^{{N\over 2}-n}$.

The preceding analysis, which is fairly standard, relied on only the 
first set (\ref{bulk/sea}) of Bethe Ansatz equations.  In order to 
compute the two-breather $S$ matrix, we also exploit the second set 
(\ref{bulk/string}) of Bethe Ansatz Eqs., which describes the centers 
of the breather strings.  Forming the product 
$\prod_{l_{1}=1}^{n_{1}}$ and taking the logarithm of both sides, we 
obtain
\be
\bar h(\lambda^{0}_{1}) = \bar J^{0}_{1} \,,
\label{BAstring}
\ee
where $\bar h(\lambda)$ is the new counting function
\be
\bar h(\lambda) = {1\over 2\pi} \left\{  N q_{n_{1}}(\lambda \,; \mu) 
- \sum_{l_{1}=1}^{n_{1}} \left[ \sum_{\beta=1}^{M^{-}_{1}}  
r_{2}(\lambda^{(n_{1},l_{1})} - \lambda_{\beta}\,; \mu) 
+ \sum_{l_{2}=1}^{n_{2}} 
q_{2}(\lambda^{(n_{1},l_{1})} - \lambda_{2}^{(n_{2},l_{2})}\,; \mu)
\right] \right\} \,,
\ee 
and $\lambda^{(n_{1},l_{1})} = \lambda + {i\over 2}( n_{1} + 1 - 2l_{1})$. 
We define the corresponding density $\bar \sigma (\lambda)$ by
\be
\bar \sigma (\lambda) = {1\over N} {d \over d \lambda} \bar 
h(\lambda) \label{barsigma} \,.
\ee 
We find
\be 
\bar \sigma (\lambda) &=& a_{n_{1}}(\lambda \,; \mu) 
- \sum_{l_{1}=1}^{n_{1}} \left\{
\int_{-\infty}^{\infty} d\lambda' \sigma(\lambda')\ 
b_{2}(\lambda^{(n_{1},l_{1})} - \lambda'\,; \mu) 
+ {1\over N} \sum_{l_{2}=1}^{n_{2}} 
a_{2}(\lambda^{(n_{1},l_{1})} - \lambda_{2}^{(n_{2},l_{2})}\,; \mu)
\right\}  \non \\ 
&=& \varepsilon_{n_{1}} (\lambda) + 
{1\over N} \sum_{l_{1}=1}^{n_{1}} \left\{ 
\sum_{\beta=1}^{2} \sum_{l_{\beta}=1}^{n_{\beta}}
(b_{2}*K_{1})(\lambda^{(n_{1},l_{1})} - 
\lambda_{\beta}^{(n_{\beta},l_{\beta})}) 
- \sum_{l_{2}=1}^{n_{2}} 
a_{2}(\lambda^{(n_{1},l_{1})} - \lambda_{2}^{(n_{2},l_{2})}\,; \mu) 
\right\} \,. \non \\ 
\label{desired}
\ee 
In passing to the second line, we have used the result (\ref{sigma1}) 
for $\sigma(\lambda)$.

\subsection{Breather bulk $S$ matrix}

We define the 
two-breather $S$ matrix $S^{(n_{1}, n_{2})}(\lambda^{0}_{1}, 
\lambda^{0}_{2})$ by the momentum quantization condition
\be
\left(e^{i p_{n_{1}}(\lambda^{0}_{1}) N}\
S^{(n_{1}, n_{2})}(\lambda^{0}_{1}, \lambda^{0}_{2}) - 1 \right) 
| \lambda^{0}_{1}, \lambda^{0}_{2} \rangle = 0 \,, 
\label{quantization} 
\ee
where the breather momentum $p_{n}(\lambda)$ is given by Eq.  
(\ref{breathermomentum}).  To compute the $S$ matrix, we use the 
identity
\be 
{1\over 2\pi} {d\over d\lambda}p_{n_{1}}(\lambda) 
+ \bar \sigma(\lambda) - \varepsilon_{n_{1}}(\lambda) 
- {1\over N} {d\over d\lambda}\bar h(\lambda) = 0 \,,
\label{key}
\ee
which immediately follows from Eqs.  (\ref{energy/momentum}) and 
(\ref{barsigma}).  Multiplying by $i 2 \pi N$, integrating with 
respect to $\lambda$ from $-\infty$ to $\lambda^{0}_{1}$, and noting 
the Bethe Ansatz Eq.  (\ref{BAstring}), we conclude that (up to a 
rapidity-independent phase factor)
\be
S^{(n_{1}, n_{2})} \sim \exp \left\{
i 2 \pi N \int_{-\infty}^{\lambda^{0}_{1}} d\lambda \  
\left( \bar \sigma (\lambda) - \varepsilon_{n_{1}} (\lambda) \right) 
\right\} \,.
\ee
Substituting our result (\ref{desired}) for $\bar \sigma (\lambda)$, 
we obtain
\be
S^{(n_{1}, n_{2})} &=& \prod_{l_{1}=1}^{n_{1}} \prod_{l_{2}=1}^{n_{2}} 
S^{(1, 1)}\left( \lambda_{1}^{(n_{1},l_{1})}
- \lambda_{2}^{(n_{2},l_{2})} \right) \non \\ 
&=& \prod_{l_{1}=1}^{n_{1}} \prod_{l_{2}=1}^{n_{2}} 
S^{(1, 1)}\left( \lambda_{1}^{0} - \lambda_{2}^{0} 
+ {i\over 2}(n_{1} - n_{2} - 2l_{1} + 2l_{2}) \right) \,,
\ee
where $S^{(1, 1)}(\lambda)$ is given by
\be
S^{(1, 1)}(\lambda) &\sim& \exp \left\{
-2\int_{0}^{\infty} {d \omega \over \omega}
\sinh \left( i \omega \lambda \right) 
{\cosh  \left( (\nu - 3) {\omega \over 2} \right) \over 
\cosh  \left( (\nu -1) {\omega \over 2} \right)} \right\} \non \\
&=& {\sinh \left( {\pi \lambda \over \nu - 1} \right)  
+ i \cos \left({\pi\over 2} ({\nu - 3\over \nu - 1}) \right) \over 
\sinh \left( {\pi \lambda \over \nu - 1} \right)  
- i \cos \left({\pi\over 2} ({\nu - 3\over \nu - 1}) \right)} \,.
\label{bulk2breather}
\ee 
This coincides with the sine-Gordon breather $S$ matrix 
\cite{ZZ},\cite{korepin}, provided that we make the identification 
$\beta^{2} = 8 \mu$ which we have already noted 
(\ref{identifications}).  The breather $S$ matrix has been obtained 
for the XXZ chain previously using the so-called physical Bethe Ansatz 
Eqs.  in \cite{KR}.

\section{Open XXZ chain}

In this Section we consider the critical open XXZ spin chain with 
boundary magnetic fields $h(\mu \,, \xi_{\pm}^{(\epsilon)})$ which are 
parallel to the symmetry axis \footnote{The dependence on $\epsilon$ 
is discussed in Appendix A.}
\be
{\cal H} =   {\epsilon\over 4} \left\{ \sum_{n=1}^{N-1} \left( 
\sigma^{x}_n \sigma^{x}_{n+1}
+ \sigma^{y}_n \sigma^{y}_{n+1} 
+ \cos \mu \ \sigma^{z}_n \sigma^{z}_{n+1} \right)  
+ h(\mu \,, \xi_{-}^{(\epsilon)}) \sigma^z_1 
+ h(\mu \,, \xi_{+}^{(\epsilon)}) \sigma^z_N 
\right\} \,, \label{open} 
\ee 
where
\be
h(\mu \,, \xi) = \sin \mu \cot (\mu \xi) \,,
\ee 
with $0 < \mu < {\pi\over 2}$ and $\epsilon = \pm 1$.  For simplicity, 
we restrict to $h(\mu \,, \xi_{\pm}^{(\epsilon)}) \le 0$.

Choosing again as the pseudovacuum the state with all spins up,
the Bethe Ansatz equations are  \cite{ABBBQ}, \cite{sklyanin}
\be 
e_{2\xi^{(\epsilon)}_{+}-1}(\lambda_{\alpha}\,; \mu)\ 
e_{2\xi^{(\epsilon)}_{-}-1}(\lambda_{\alpha}\,; \mu)\
e_{1}(\lambda_{\alpha}\,; \mu)^{2N} &=& 
\prod_{\scriptstyle{\beta=1}\atop \scriptstyle{\beta \ne \alpha}}^M 
e_{2}(\lambda_{\alpha}-\lambda_{\beta}\,; \mu)\ 
e_{2}(\lambda_{\alpha}+\lambda_{\beta}\,; \mu)  \,, \non \\ 
& & \alpha = 1 \,, \cdots \,, M \,.
\label{BAEopen}
\ee
To streamline the notation, we shall often suppress the superscript 
$(\epsilon)$ and thus write the boundary parameters 
$\xi^{(\epsilon)}_{\pm}$ as $\xi_{\pm}$.

The energy is given by Eq.  (\ref{energy}) (plus terms that are 
independent of $\{ \lambda_{\alpha} \}$) and the $S^{z}$ eigenvalue is 
again given by Eq.  (\ref{spin}).  The requirement that Bethe Ansatz 
solutions correspond to independent Bethe Ansatz states leads to the 
restriction (see \cite{FS},\cite{GMN} and references therein)
\be
Re \left( \lambda_{\alpha} \right) \ge  0  \,, \qquad 
\lambda_{\alpha} \ne 0 \,, \infty  \,.
\label{lambdarestriction}
\ee

In addition to having the well-known ``bulk'' string solutions, the 
Bethe Ansatz Eqs.  for the open chain also have ``boundary'' string 
solutions \cite{SS}.  In particular, there are boundary 1-strings 
$\lambda = \pm i \left( \xi_{\pm} - {1\over 2} \right)$ for ${1\over 
2} - {\nu\over 2} < \xi_{\pm} < {1\over 2}$ $(\mbox{mod } \nu)$.  (See 
Appendix B.) For simplicity, we shall restrict $\xi_{\pm}$ so that 
such strings are absent, namely,
\be
-{\nu\over 2} \le  \xi_{\pm} < {1\over 2} - {\nu\over 2} \qquad 
(\mbox{mod } \nu) \,.
\label{xirestriction}
\ee 
The lower bound comes from the restriction 
$h(\mu \,, \xi_{\pm}) \le 0$.

\subsection{One-breather state}

We consider again the attractive case $\epsilon = -1$.  For values of 
$\xi_{\pm}$ in the range (\ref{xirestriction}), there are no boundary 
strings; and hence, the ground state is a sea of negative-parity 
1-strings, as already discussed for the closed chain in Section 2.1.

In order to compute the breather boundary $S$ matrix, we consider the 
Bethe Ansatz state consisting of one breather 
$\lambda_{0}^{(n ,l)}$ in the sea. The corresponding Bethe Ansatz 
Eqs. read
\be 
\lefteqn{g_{2\xi_{+}-1}(\lambda_{\alpha}\,; \mu)\ 
g_{2\xi_{-}-1}(\lambda_{\alpha}\,; \mu)\
e_{1}(\lambda_{\alpha}\,; \mu)\
g_{1}(\lambda_{\alpha}\,; \mu)^{2N+1}} \non \\
&=& -\prod_{\beta=1}^{M^{-}_{1}} 
e_{2}(\lambda_{\alpha}-\lambda_{\beta}\,; \mu)\ 
e_{2}(\lambda_{\alpha}+\lambda_{\beta}\,; \mu) 
\prod_{l=1}^{n}
g_{2}(\lambda_{\alpha}- \lambda_{0}^{(n,l)}
\,; \mu)\
g_{2}(\lambda_{\alpha}+ \lambda_{0}^{(n,l)}
\,; \mu) \,, \non \\
& &  \qquad \qquad \qquad \alpha = 1 \,, \cdots \,, M^{-}_{1} \,, 
\label{boundary/sea} 
\ee
\be 
\lefteqn{e_{2\xi_{+}-1}(\lambda_{0}^{(n,l)}\,; \mu)\
e_{2\xi_{-}-1}(\lambda_{0}^{(n,l)}\,; \mu)\
e_{2}(2\lambda_{0}^{(n,l)}\,; \mu)\
e_{1}(\lambda_{0}^{(n,l)}\,; \mu)^{2N}} \non \\
&=& -\prod_{\beta=1}^{M^{-}_{1}} 
g_{2}(\lambda_{0}^{(n,l)} -\lambda_{\beta}\,; \mu)\
g_{2}(\lambda_{0}^{(n,l)} +\lambda_{\beta}\,; \mu)
\prod_{k=1}^{n} 
e_{2}(\lambda_{0}^{(n,l)} - \lambda_{0}^{(n,k)}\,; \mu)\
e_{2}(\lambda_{0}^{(n,l)} + \lambda_{0}^{(n,k)}\,; \mu)\
\,, \non \\  
& & \qquad \qquad \qquad \qquad l = 1 \,, \cdots \,, n \,. 
\label{boundary/string}
\ee

The first set of equations (\ref{boundary/sea}) leads to the counting 
function 
\be
\lefteqn{h(\lambda) = -{1\over 2\pi} \Bigg\{  
(2 N+1) r_{1}(\lambda\,; \mu) 
+ q_{1}(\lambda\,; \mu) + r_{2\xi_{+}-1}(\lambda\,; \mu)
+ r_{2\xi_{-}-1}(\lambda\,; \mu)} \non \\ 
& & -\sum_{\beta=1}^{M^{-}_{1}} \left[
q_{2}(\lambda - \lambda_{\beta}\,; \mu) 
+ q_{2}(\lambda + \lambda_{\beta}\,; \mu) \right] 
-  \sum_{l=1}^{n} \left[
r_{2}(\lambda - \lambda_{0}^{(n,l)}\,; \mu)
+ r_{2}(\lambda + \lambda_{0}^{(n,l)}\,; \mu) \right]
\Bigg\} \,. 
\ee 
We define the corresponding density $\sigma(\lambda)$ as before 
(\ref{sigma}).  The restriction (\ref{lambdarestriction}) on the 
Bethe Ansatz roots implies that we must pass from sums to integrals 
using \cite{FS},\cite{GMN}
\be 
{1\over N}\sum_{\alpha=1}^{M^{-}_{1}} g(\lambda_\alpha)
= \int_0^{\infty} d\lambda' \  \sigma(\lambda')\ g(\lambda') 
- {1\over 2N} g(0) 
\label{euler}
\ee
(plus terms that are of higher order in $1/N$), where $g(\lambda)$ is 
an arbitrary function.  We arrive in this way at the following linear 
integral equation for $\sigma(\lambda)$:
\be
\lefteqn{\sigma(\lambda) = -2 b_{1}(\lambda \,; \mu) 
+ \int_{0}^{\infty} d\lambda' \ \left[
a_{2}(\lambda - \lambda'\,; \mu) + a_{2}(\lambda + \lambda'\,; \mu) 
\right] \sigma(\lambda')} \non \\
&-&{1\over N} \left[ a_{1}(\lambda \,; \mu) + a_{2}(\lambda \,; \mu) 
+ b_{1}(\lambda \,; \mu) + b_{2\xi_{+}-1}(\lambda \,; \mu) 
+ b_{2\xi_{-}-1}(\lambda \,; \mu) \right] \non \\ 
&+& {1\over N}\sum_{l=1}^{n} \left[
b_{2}(\lambda - \lambda_{0}^{(n,l)}\,; \mu)
+ b_{2}(\lambda + \lambda_{0}^{(n,l)}\,; \mu) \right] \,, \quad 
\lambda > 0 \,.
\ee 
Finally, defining the symmetric density $\sigma_s(\lambda)$ by 
\be
\sigma_s(\lambda) = \left\{  
\begin{array}{lll} 
         \sigma(\lambda)  & \lambda > 0 \\
         \sigma(-\lambda) & \lambda < 0  
\end{array} \right. \,,
\label{symmetric}  
\ee
we see that it is given by
\be 
\sigma_{s}(\lambda) &=& 2 s (\lambda) 
- {1\over N} \sum_{l=1}^{n} \left( 
K_{1}(\lambda - \lambda_{0}^{(n,l)}) +
K_{1}(\lambda + \lambda_{0}^{(n,l)}) \right) \non \\
&+&{1\over N} \left( R * \left( 
a_{1} + a_{2} + b_{1} +  b_{2\xi_{+}-1} + b_{2\xi_{-}-1} \right) \right) 
(\lambda) \,,
\label{sigmasbound}
\ee 
where the Fourier transforms of $R(\lambda)$ and $K_{n}(\lambda)$ are 
given by
\be
\hat R(\omega) = {1\over -1 + \hat a_{2}(\omega \,; \mu) } \,,
\ee 
and Eq. (\ref{Kn}), respectively.

We turn now to the second set of Bethe Ansatz equations 
(\ref{boundary/string}).  Forming the product $\prod_{l=1}^{n}$ and 
taking the logarithm of both sides, we obtain the counting function
\be
\bar h(\lambda) &=& {1\over 2\pi} \Bigg\{  2 N q_{n}(\lambda\,; \mu) 
+ \sum_{l=1}^{n} \left[ q_{2\xi_{+}-1}(\lambda^{(n,l)} \,; \mu)
+ q_{2\xi_{-}-1}(\lambda^{(n,l)} \,; \mu) 
+ q_{2}(2\lambda^{(n,l)} \,; \mu) \right] \non \\ 
&-&  \sum_{l=1}^{n} \sum_{\beta=1}^{M^{-}_{1}} \left[
r_{2}(\lambda^{(n,l)} - \lambda_{\beta}\,; \mu) 
+ r_{2}(\lambda^{(n,l)} + \lambda_{\beta}\,; \mu) \right] \non \\
&-&  \sum_{l,k=1}^{n} \left[
q_{2}(\lambda^{(n,l)} - \lambda^{(n,k)}\,; \mu)
+ q_{2}(\lambda^{(n,l)} + \lambda^{(n,k)}\,; \mu) \right]
\Bigg\} \,,
\ee 
where $\lambda^{(n,l)} = \lambda + {i\over 2}( n + 1 - 2l)$.
We find that the corresponding density $\bar \sigma (\lambda)$, defined as in 
Eq. (\ref{barsigma}), is given by
\be
{\bar \sigma} (\lambda) = 2 \varepsilon_{n}(\lambda)
+ {1\over N} \sum_{l=1}^{n}  \Bigg\{
b_{2} (\lambda^{(n,l)} \,; \mu) +  a_{2\xi_{+}-1}(\lambda^{(n,l)} \,; \mu)
+ a_{2\xi_{-}-1}(\lambda^{(n,l)} \,; \mu) 
+ 2 a_{2}(2 \lambda^{(n,l)} \,; \mu) \non \\
- \left( K_{1} *
\left( a_{1} + a_{2} + b_{1} +  b_{2\xi_{+}-1} + b_{2\xi_{-}-1} 
\right) \right) (\lambda^{(n,l)}) \Bigg\} \non \\
+ {1\over N} \sum_{l,k=1}^{n} \Bigg\{
\left( K_{1} * b_{2} \right)(\lambda^{(n,l)} - \lambda_{0}^{(n,k)})
+ \left( K_{1} * b_{2} \right)(\lambda^{(n,l)} + \lambda_{0}^{(n,k)})
- 2 a_{2} (\lambda^{(n,l)} + \lambda^{(n,k)}\,; \mu)
 \Bigg\} \non  \\
\,.
\label{desiredbreather}
\ee 
In obtaining this result, we have again used (\ref{euler}) to pass 
from a sum to an integral, and then we have used our result 
(\ref{sigmasbound}) for the density $\sigma_{s}(\lambda)$

\subsection{Breather boundary $S$ matrix}

We define the boundary $S$ matrix 
$S^{(n)}(\lambda_{0} \,, \xi)$ for the breather
$\lambda^{(n,l)}_{0} = \lambda_{0} + {i\over 2}( n + 1 - 2l) \,, $
$l=1 \,, \ldots \,, n$ by the quantization condition
\be
\left(e^{i 2 p_{n}(\lambda_{0}) N}\
S^{(n)}(\lambda_{0} \,, \xi_{-})\ 
S^{(n)}(\lambda_{0} \,, \xi_{+})
 - 1 \right) 
| \lambda_{0} \rangle = 0 \,, 
\label{boundaryquantization} 
\ee
where $p_{n}(\lambda)$ is given by Eq.  (\ref{breathermomentum}).  To 
compute the $S$ matrix, we make use of the identity 
\be 
{1\over \pi} {d\over d\lambda}p_{n}(\lambda) 
+ \bar \sigma(\lambda) - 2 \varepsilon_{n}(\lambda) 
- {1\over N} {d\over d\lambda}\bar h(\lambda) = 0 \,,
\ee
which is similar to (\ref{key}), and obtain (up to a 
rapidity-independent phase factor)
\be
S^{(n)}(\lambda_{0} \,, \xi_{-})\ 
S^{(n)}(\lambda_{0} \,, \xi_{+})
\sim \exp \left\{
i 2 \pi N \int_{0}^{\lambda_{0}} d\lambda \  
\left( \bar \sigma (\lambda) - 2\varepsilon_{n} (\lambda) \right) 
\right\} \,.
\ee
Substituting our result (\ref{desiredbreather}) for $\bar \sigma 
(\lambda)$, we obtain \footnote{We assume that the terms which 
do not depend on $\xi_{\pm}$ contribute equally to 
$S^{(n)}(\lambda_{0} \,, \xi_{-})$ and $S^{(n)}(\lambda_{0} \,, \xi_{+})$.}
\be
S^{(n)}(\lambda_{0} \,, \xi_{\pm}) = S^{(n)}_{0}(\lambda_{0})\  
S^{(n)}_{1}(\lambda_{0} \,, \xi_{\pm}) \,,
\label{breatherres1}
\ee
where
\be 
S^{(n)}_{0}(\lambda_{0}) = \prod_{l=1}^{n} 
S^{(1)}_{0}(\lambda_{0}^{(n,l)}) \
\prod_{l < k}^{n} S^{(1,1)}(\lambda_{0}^{(n,l)} + \lambda_{0}^{(n,k)}) 
\,,
\ee
and
\be
S^{(1)}_{0}(\lambda_{0}) &=& \exp \left\{
-2\int_{0}^{\infty}{d\omega\over \omega} 
\sinh ( 2 i \omega \lambda_{0} ) 
{\cosh \left( {\omega\over 2} \right) \cosh \left( {\nu \omega\over 2} \right) 
\over 
\cosh \left( (\nu - 1){\omega \over 2} \right)
\cosh \left( (\nu - 1)\omega \right)} \right\} \non  \\
&=&
{\sin \left( {i \pi \lambda_{0}\over 2 (\nu - 1)}
- {\pi\over 4 (\nu - 1)} + {\pi\over 2}  \right) 
\sin \left( {i \pi \lambda_{0}\over 2 (\nu - 1)}
+{\pi\over 4 (\nu - 1)} - {\pi\over 4} \right) 
\sin \left( {i \pi \lambda_{0}\over 2 (\nu - 1)} 
- {\pi\over 4} \right) \over 
\sin \left( {i \pi \lambda_{0}\over 2 (\nu - 1)} 
+ {\pi\over 4 (\nu - 1)} + {\pi\over 2} \right) 
\sin \left( {i \pi \lambda_{0}\over 2 (\nu - 1)} 
-{\pi\over 4 (\nu - 1)} + {\pi\over 4} \right)
\sin \left( {i \pi \lambda_{0}\over 2 (\nu - 1)} 
+ {\pi\over 4} \right) } \,.
\label{breatherres2}
\ee
We recall that $S^{(1,1)}(\lambda)$ is the bulk two-breather $S$ 
matrix (\ref{bulk2breather}), and that $\nu = {\pi\over \mu}$. 
Moreover, 
\be
S^{(n)}_{1}(\lambda_{0} \,, \xi) = \prod_{l=1}^{n} 
S^{(1)}_{1}(\lambda_{0}^{(n,l)} \,, \xi) \,,
\ee
where
\be 
S^{(1)}_{1}(\lambda_{0} \,, \xi) &=& 
\exp \left\{
2\int_{0}^{\infty}{d\omega\over \omega} 
\sinh ( 2 i \omega \lambda_{0} )
{\cosh \left( (\nu - 2\xi ) \omega \right)
\over \cosh \left( (\nu - 1) \omega  \right)} \right\} \non  \\
&=& 
{-\sin \left( {i \pi \lambda_{0}\over \nu - 1} \right)
- \cos \left( {\pi (\nu - 2 \xi)\over 2(\nu - 1)} \right) \over 
\sin \left( {i \pi \lambda_{0}\over \nu - 1} \right)
- \cos \left( {\pi (\nu - 2 \xi)\over 2(\nu - 1)} \right)} 
\,.
\label{breatherres3}
\ee 
It can be shown that this result 
agrees with Ghoshal's bootstrap result \cite{ghoshal} for the breather 
boundary $S$ matrix for the boundary sine-Gordon model with ``fixed'' 
boundary conditions, provided that we make the identification of bulk 
coupling constants $\beta^{2} = 8 \mu$ (see Eq.  (\ref{identifications})), 
as well as the identification of boundary 
parameters \footnote{We 
denote the Ghoshal-Zamolodchikov \cite{GZ},\cite{ghoshal} boundary 
parameter $\xi$ by $x$, in order to distinguish it from our boundary 
parameter $\xi$.  We recall that Ghoshal-Zamolodchikov identify as
``fixed'' boundary conditions their case $k=0$.  Moreover, 
in the attractive case, their bulk coupling constant $\lambda = 
{8\pi\over \beta^2} - 1$ is related to our coupling constant $\nu = 
{\pi\over \mu}$  by $\lambda = \nu - 1$; and their 
rapidity variable $\theta$ is related to our variable $\lambda_{0}$ by 
$\theta= {\pi \lambda_{0}\over \nu - 1}$. \label{notations}}
\be
x= {\pi\over 2}\left( 2 \xi^{(-1)} - \nu \right) \,.
\label{boundaryidentifications}
\ee 
In this formula we have restored the superscript $(\epsilon)$ 
on the boundary parameter $\xi$.

\begin{figure}[htb]
	\centering
	\epsfxsize=.25\textwidth\epsfbox{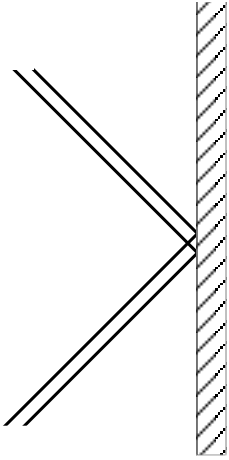}
	\caption[xxx]{\parbox[t]{.7\textwidth}{
	Scattering of a 2-breather from the boundary.}
	}
	\label{fig1}
\end{figure}

We remark that the appearance of the bulk $S$ matrix in our 
expression for the boundary $S$ matrix can be readily understood from 
the fact that an $n$-breather can be regarded as a bound state of $n$ 
1-breathers, which scatter among themselves upon reflection from the 
boundary. This is illustrated in Figure \ref{fig1} for the case $n=2$.
A single line represents a 1-breather, and so the 2-breather is 
represented by a double line.

\subsection{Soliton boundary $S$ matrix}

Although the main focus of this paper is on breather $S$ matrices, we 
compute here the soliton boundary $S$ matrix in order to further check 
the identification (\ref{boundaryidentifications}) of the boundary 
parameters.

\subsubsection{Attractive case ($\epsilon = -1$)}

We consider first the attractive case $\epsilon = -1$, with 
$\xi_{\pm}$ in the range (\ref{xirestriction}), and so the ground 
state is a sea of negative-parity 1-strings.  Following 
\cite{GMN},\cite{DMN}, we consider the Bethe Ansatz state consisting 
of one hole with rapidity $\tilde \lambda$ in the sea,
which has $S^z = +{1\over 2}$. The counting function is
\be
h(\lambda) &=& -{1\over 2\pi} \Bigg\{  
(2 N+1) r_{1}(\lambda\,; \mu) 
+ q_{1}(\lambda\,; \mu) + r_{2\xi_{+}-1}(\lambda\,; \mu)
+ r_{2\xi_{-}-1}(\lambda\,; \mu) \non \\ 
&-& \sum_{\beta=1}^{M^{-}_{1}} \left[
q_{2}(\lambda - \lambda_{\beta}\,; \mu) 
+ q_{2}(\lambda + \lambda_{\beta}\,; \mu) \right] 
\Bigg\} \,,
\ee 
which leads to the density $\sigma_{s}(\lambda)$ whose Fourier 
transform is given by \footnote{Due to the presence of the hole,
the prescription (\ref{euler}) for passing from sums to integrals has
the additional term $-{1\over N} g(\tilde \lambda)$ on the 
right-hand-side.}
\be 
\hat \sigma_{s}(\omega) &=& 2 \hat s (\omega) 
- {1\over N}\hat J(\omega) \left(
e^{i \omega  \tilde\lambda} 
+ e^{-i \omega \tilde\lambda} \right) 
+{1\over N \left(-1 + \hat a_{2}(\omega \,; \mu) \right)} 
\bigg[ \hat a_{1}(\omega \,; \mu) \non \\
&+& \hat a_{2}(\omega \,; \mu)
+ \hat b_{1}(\omega \,; \mu) + \hat b_{2\xi_{+}-1}(\omega \,; \mu)
+ \hat b_{2\xi_{-}-1}(\omega \,; \mu) \bigg] \,,
\label{sigmasboundsoliton}
\ee 
where $\hat J(\omega)$ is defined by 
\be 
\hat J(\omega) = {\hat a_{2}(\omega \,; \mu) \over 
1 - \hat a_{2}(\omega \,; \mu)} = 
{\sinh \left( (\nu - 2) {\omega \over 2} \right)  \over 
2 \sinh \left( {\omega \over 2} \right) 
\cosh  \left( (\nu -1) {\omega \over 2} \right)} \,.
\label{J}
\ee 

We define the boundary $S$ matrix $S(\tilde\lambda \,, \xi_{\pm})$ for 
the soliton by the quantization condition
\be
\left(e^{i 2 p(\tilde\lambda) N}\
S(\tilde\lambda \,, \xi_{-})\ 
S(\tilde\lambda \,, \xi_{+})
 - 1 \right) 
| \tilde\lambda \rangle = 0 \,, 
\label{quantizationsoliton} 
\ee
where $p(\lambda)$ is given by
\be
p(\lambda) = -\left( J * r_{1} \right) (\lambda) - 
r_{1}(\lambda \,; \mu)  \,.
\ee
The boundary $S$ matrix has the diagonal form
\be
S(\tilde\lambda \,, \xi_\pm) = 
\left( \begin{array}{ll}
      \alpha(\tilde\lambda \,, \xi_\pm) &0  \\
      0  & \beta(\tilde\lambda \,, \xi_\pm)  \end{array}\right) 
\,. \label{form} 
\ee 
The matrix elements 
$\alpha(\tilde\lambda \,, \xi)$ and $\beta(\tilde\lambda \,, \xi)$ 
are the boundary scattering amplitudes for one-hole states
with $S^z = +{1\over 2}$ and $S^z = -{1\over 2}$, respectively.
We compute these matrix elements with the help of the identity 
\be 
{1\over \pi} {d\over d\lambda}p(\lambda) 
+ \sigma_{s}(\lambda) - 2 s(\lambda) 
- {1\over N} {d\over d\lambda}h(\lambda) = 0 \,.
\ee

We first compute $\alpha(\tilde\lambda \,, \xi)$. We have
(up to a rapidity-independent phase factor)
\be
\alpha(\tilde\lambda \,, \xi_{+})\ \alpha(\tilde\lambda \,, \xi_{-})  
\sim \exp \left\{ i 2\pi N \int_{0}^{\tilde\lambda}
\left( \sigma_{s}(\lambda) - 2 s(\lambda) \right) d\lambda
\right\} \,.
\ee 
Substituting the result (\ref{sigmasboundsoliton}) for the root 
density and performing some algebra, we obtain 
\be
\alpha(\tilde\lambda \,, \xi) &\sim& \exp \Bigg\{ 
-2 \int_{0}^{\infty}{d\omega\over \omega} 
\sinh \left( 2 i \omega \tilde\lambda \right) \bigg[
{\sinh \left( 3(\nu - 1){\omega\over 2} \right)
\sinh \left( (\nu - 2){\omega\over 2} \right) \over 
\sinh \left( {\omega\over 2} \right)
\sinh \left( 2(\nu - 1)\omega \right)} \non \\
&-& {\sinh \left( (2 \xi - 1) \omega \right) \over
2 \sinh \omega \cosh \left( (\nu - 1) \omega \right)} \bigg] \Bigg\} 
\non \\
&=& {1\over \pi} \cosh [ \pi ( \tilde \lambda + {i\over 2}(2 \xi - 
\nu) ) ]\ 
S_{0}(\tilde\lambda)\ S_{1}(\tilde\lambda \,, \xi)
\,, \label{alphaa}
\ee
where
\be
S_{0}(\tilde\lambda) &=& \prod_{n=0}^{\infty} \Bigg\{
{\Gamma \left( -2i\tilde\lambda + (\nu-1)(4n + 3) + 1 \right)
\Gamma \left( -2i\tilde\lambda + (\nu-1)(4n + 1) \right) \over 
\Gamma \left( 2i\tilde\lambda + (\nu-1)(4n + 3) + 1 \right)
\Gamma \left( 2i\tilde\lambda + (\nu-1)(4n + 1) \right)} \non \\
&\times& {\Gamma \left( 2i\tilde\lambda + 4(\nu-1)n + 1 \right)
\Gamma \left( 2i\tilde\lambda + 4(\nu-1)(n + 1) \right) \over
\Gamma \left( -2i\tilde\lambda + 4(\nu-1)n + 1 \right)
\Gamma \left( -2i\tilde\lambda + 4(\nu-1)(n + 1) \right)} \,,
\label{s0a}
\ee
and
\be
S_{1}(\tilde\lambda \,, \xi) &=& 
\prod_{n=0}^{\infty}\Bigg\{
{\Gamma\left(-i\tilde\lambda + 2(\nu-1)n-{1\over 2}(2\xi - \nu -1)\right)
 \Gamma\left(-i\tilde\lambda + 2(\nu-1)n+{1\over 2}(2\xi - \nu +1)\right)
\over
\Gamma\left(i\tilde\lambda + (\nu-1)(2n+2)-{1\over 2}(2\xi - \nu -1)\right)
\Gamma\left(i\tilde\lambda + (\nu-1)(2n+2)+{1\over 2}(2\xi - \nu +1)\right)}
\non  \\
&\times&
{\Gamma\left(i\tilde\lambda + (\nu-1)(2n+1)-{1\over 2}(2\xi - \nu -1)\right)
\Gamma\left(i\tilde\lambda + (\nu-1)(2n+1)+{1\over 2}(2\xi - \nu +1)\right)
\over
\Gamma\left(-i\tilde\lambda + (\nu-1)(2n+1)-{1\over 2}(2\xi - \nu -1)\right)
\Gamma\left(-i\tilde\lambda + (\nu-1)(2n+1)+{1\over 2}(2\xi - \nu +1)\right)}
\Bigg\} \,. \non  \\
\label{s1a}
\ee 

In order to compute $\beta(\tilde\lambda \,, \xi)$, 
we must consider a one-hole state with $S^z = -{1\over 2}$. As explained 
in \cite{GMN},\cite{DMN}, this can be achieved by working instead with the 
the pseudovacuum with all spins down, in which case the Bethe Ansatz 
Eqs. are given by (\ref{BAEopen}) with the replacement 
$\xi_\pm \rightarrow -\xi_\pm$. 
The corresponding density $\hat\sigma'_{s}(\omega)$ is given by 
Eq. (\ref{sigmasboundsoliton}) with the replacement
$\hat b_{2\xi_{\pm}-1}(\omega \,; \mu) \rightarrow 
-\hat b_{2\xi_{\pm}+1}(\omega \,; \mu)$. 
We find
\be
\sigma'_{s}(\lambda) - \sigma_{s}(\lambda) = {1\over N} \left(
b_{2\xi_{-} - \nu}(\lambda \,; \pi) + 
b_{2\xi_{+} - \nu}(\lambda \,; \pi) \right) \,.
\ee 
In obtaining this result, we have noted that for $\xi_{\pm}$ in the 
range (\ref{xirestriction}), the quantities $\hat 
b_{2\xi_{\pm}-1}(\omega \,; \mu)$ and $\hat b_{2\xi_{\pm}+1}(\omega 
\,; \mu)$ are given by Eqs.  (\ref{fourier2}) and (\ref{fourier3}), 
respectively. Since
\be
{\beta(\tilde\lambda \,, \xi_{+})\ \beta(\tilde\lambda \,, \xi_{-})\over
\alpha(\tilde\lambda \,, \xi_{+})\ \alpha(\tilde\lambda \,, \xi_{-})}
= \exp \left\{ i 2\pi N \int_{0}^{\tilde\lambda}
\left( \sigma'_{s}(\lambda) - \sigma_{s}(\lambda) \right) d\lambda
\right\} \,,
\ee 
we conclude that
\be
{\beta(\tilde\lambda \,, \xi)\over \alpha(\tilde\lambda \,, \xi)}
= {\cosh \left[ \pi \left( \tilde \lambda 
- {i\over 2}(2 \xi - \nu) \right) \right]
\over
\cosh \left[ \pi \left( \tilde \lambda 
+ {i\over 2}(2 \xi - \nu) \right) \right]}
\,.
\label{betaa}
\ee

The soliton boundary $S$ matrix (\ref{form}), 
(\ref{alphaa})-(\ref{s1a}),(\ref{betaa}) agrees with the bootstrap 
result of Ghoshal-Zamolodchikov \cite{GZ} for the boundary sine-Gordon 
model with ``fixed'' boundary conditions, provided the  
identification of boundary parameters (\ref{boundaryidentifications}) 
is again made. Fendley-Saleur \cite{FS} find a similar identification.

\subsubsection{Repulsive case ($\epsilon = +1$)}

We consider finally the repulsive case $\epsilon = +1$, where there 
are no breathers.  Here the ground state corresponds to a sea of {\it 
positive} parity 1-strings, i.e., real solutions of the Bethe Ansatz 
equations.  For the Bethe Ansatz state with one hole of rapidity 
$\tilde \lambda$ in the sea, the counting function is
\be
h(\lambda) &=& {1\over 2\pi} \Bigg\{  
(2 N+1) q_{1}(\lambda\,; \mu) 
+ r_{1}(\lambda\,; \mu) + q_{2\xi_{+}-1}(\lambda\,; \mu)
+ q_{2\xi_{-}-1}(\lambda\,; \mu) \non \\ 
&-& \sum_{\beta=1}^{M^{-}_{1}} \left[
q_{2}(\lambda - \lambda_{\beta}\,; \mu) 
+ q_{2}(\lambda + \lambda_{\beta}\,; \mu) \right] 
\Bigg\} \,,
\ee 
and the density $\sigma_{s}(\lambda)$ is given by
\be 
\hat \sigma_{s}(\omega) &=& 2 \hat s (\omega) 
+ {1\over N}\hat J(\omega) \left(
e^{i \omega  \tilde\lambda} 
+ e^{-i \omega \tilde\lambda} \right) 
+{1\over N \left(1 + \hat a_{2}(\omega \,; \mu) \right)} 
\bigg[ \hat a_{1}(\omega \,; \mu) \non \\
&+& \hat a_{2}(\omega \,; \mu)
+ \hat b_{1}(\omega \,; \mu) + \hat a_{2\xi_{+}-1}(\omega \,; \mu)
+ \hat a_{2\xi_{-}-1}(\omega \,; \mu) \bigg] \,,
\ee 
where now
\be
\hat s(\omega) &=& {\hat a_{1}(\omega\,; \mu)
\over 1 + \hat a_{2}(\omega\,; \mu)} = 
{1\over 2 \cosh \left( {\omega\over 2} \right)} \,, \non \\
\hat J(\omega) &=& 
{\hat a_{2}(\omega\,; \mu)\over 1 + \hat a_{2}(\omega\,; \mu)} =
{\sinh \left( (\nu  - 2) {\omega \over 2}) \right) \over
2 \sinh \left( (\nu - 1){\omega \over 2} \right)
\cosh \left( {\omega \over 2} \right)} \,.
\ee
Moreover, in the repulsive case,
\be
p(\lambda) = -\left( J * q_{1} \right) (\lambda) + 
q_{1}(\lambda \,; \mu)  \,.
\ee 
Proceeding as in the attractive case, we find that the soliton boundary $S$ 
matrix has the form (\ref{form}) with matrix elements
\be
\alpha(\tilde\lambda \,, \xi) &\sim& \exp \Bigg\{ 
2 \int_{0}^{\infty}{d\omega\over \omega} 
\sinh \left( 2 i \omega \tilde\lambda \right) \bigg[
{\sinh \left( {3\omega\over 2} \right)
\sinh \left( (\nu - 2){\omega\over 2} \right) \over 
\sinh \left( 2\omega \right)
\sinh \left( (\nu - 1){\omega \over 2} \right)} \non \\
&+& {\sinh \left( (\nu - 2 \xi + 1) \omega \right) \over
2 \sinh\left( (\nu - 1) \omega \right) \cosh \omega} \bigg] \Bigg\}
\non \\
&=& {1\over \pi} \cosh [ {\pi\over \nu - 1} ( \tilde \lambda 
+ {i\over 2}(\nu - 2 \xi) ) ]\ 
S_{0}(\tilde\lambda)\ S_{1}(\tilde\lambda \,, \xi)
\,, \label{alphar}
\ee
where
\be
S_{0}(\tilde\lambda) &=& \prod_{n=0}^{\infty} \Bigg\{
{\Gamma \left( {1\over \nu -1}(-2i\tilde\lambda + 4n + 3) + 1 \right)
 \Gamma \left( {1\over \nu -1}(-2i\tilde\lambda + 4n + 1) \right) \over 
 \Gamma \left( {1\over \nu -1}(2i\tilde\lambda + 4n + 3) + 1 \right)
 \Gamma \left( {1\over \nu -1}(2i\tilde\lambda + 4n + 1) \right)} 
\non \\
&\times& 
{\Gamma \left( {1\over \nu -1}(2i\tilde\lambda + 4n) + 1 \right)
 \Gamma \left( {1\over \nu -1}(2i\tilde\lambda + 4n + 4) \right) \over
 \Gamma \left( {1\over \nu -1}(-2i\tilde\lambda + 4n) + 1 \right)
 \Gamma \left( {1\over \nu -1}(-2i\tilde\lambda + 4n + 4) \right)} \,,
\label{s0r}
\ee
\be
S_{1}(\tilde\lambda \,, \xi) &=& 
\prod_{n=0}^{\infty}\Bigg\{
{\Gamma\left( {1\over \nu -1}(-i\tilde\lambda + 2n 
-{1\over 2}(\nu - 2\xi )) +{1\over 2} \right)
 \Gamma\left( {1\over \nu -1}(-i\tilde\lambda + 2n 
+{1\over 2}(\nu - 2\xi )) +{1\over 2} \right)
\over
\Gamma\left( {1\over \nu -1}(i\tilde\lambda + 2n+2
-{1\over 2}(\nu - 2\xi)) +{1\over 2} \right)
\Gamma\left( {1\over \nu -1}(i\tilde\lambda + 2n+2
+{1\over 2}(\nu - 2\xi)) +{1\over 2} \right)}
\non  \\
&\times&
{\Gamma\left( {1\over \nu -1}(i\tilde\lambda + 2n+1
-{1\over 2}(\nu - 2\xi)) +{1\over 2} \right)
 \Gamma\left( {1\over \nu -1}(i\tilde\lambda + 2n+1
+{1\over 2}(\nu - 2\xi)) +{1\over 2} \right)
\over
\Gamma\left( {1\over \nu -1}(-i\tilde\lambda + 2n+1
-{1\over 2}(\nu - 2\xi)) +{1\over 2} \right)
\Gamma\left( {1\over \nu -1}(-i\tilde\lambda + 2n+1
+{1\over 2}(\nu - 2\xi)) +{1\over 2} \right)}
\Bigg\} \,, \non  \\
\label{s1r}
\ee 
and
\be
{\beta(\tilde\lambda \,, \xi)\over \alpha(\tilde\lambda \,, \xi)}
= {\cosh \left[ {\pi\over \nu - 1} \left( \tilde \lambda 
- {i\over 2}(\nu - 2 \xi) \right) \right]
\over
\cosh \left[ {\pi\over \nu - 1} \left( \tilde \lambda 
+ {i\over 2}(\nu - 2 \xi) \right) \right]}
\,.
\label{betar}
\ee
Comparing with Ghoshal-Zamolodchikov \cite{GZ}, we obtain the 
following identification of boundary parameters \footnote{In the 
repulsive case, the Ghoshal-Zamolodchikov bulk coupling constant 
$\lambda$ is related to our coupling constant $\nu$ by $\lambda = 
{1\over \nu - 1}$; and their rapidity variable $\theta$ is related to 
our variable $\tilde\lambda$ by $\theta = \pi \tilde\lambda$.}
\be
x= {\pi\left( \nu - 2 \xi^{(+1)} \right)\over 2 (\nu -1)}
\,.
\ee 
The same identification was found in \cite{FS}. We remark that, 
by setting
\be
\mu' = \pi - \mu \,, \qquad \xi' = - {\xi^{(-1)}\over \nu - 1} \,,
\ee
as in Appendix A, the formula (\ref{boundaryidentifications}) for $x$ 
in the attractive case can be recast in the similar form
\be
x= {\pi\left( -\nu' - 2 \xi' \right)\over 2 (\nu' -1)} \,,
\ee 
where $\nu' = {\pi\over \mu'}$.

\section{Discussion}

We have formulated a systematic Bethe-Ansatz approach for computing 
breather $S$ matrices for integrable quantum spin chains.  We have 
used this approach to calculate the breather boundary $S$ matrix for 
the open XXZ spin chain with diagonal boundary fields. We have also 
directly computed the soliton boundary $S$ matrix in the critical regime.

Let us briefly compare our approach with that of other authors.  Our 
approach is essentially a systematization of Korepin's \cite{korepin} 
analysis of the massive Thirring model.  Key elements of our approach 
are the exploitation of the ``second'' set of Bethe Ansatz Eqs.  
(\ref{bulk/string}) which describes the centers of the breather 
strings; and the use of the identity (\ref{key}).  An 
analogous identity for holes was used by Andrei and Destri 
\cite{AD} to compute soliton $S$ matrices.  Fendley and Saleur 
\cite{FS} study boundary $S$ matrices of the XXZ chain using an 
alternative approach based on the model's physical Bethe Ansatz 
equations \cite{KR}.  The identification of boundary 
$S$ matrices from the physical Bethe Ansatz Eqs.  is not 
straightforward, especially in the repulsive case. Finally, we remark 
that the vertex-operator approach \cite{kyoto} has so far been 
restricted in applicability to the noncritical regime.

While we have focused on the XXZ chain for simplicity, we expect that 
the same methods should be applicable to other models.  Indeed, 
boundary $S$ matrices for the critical $A_{\n-1}^{(1)}$ open spin chain 
with diagonal boundary fields can be computed in this way \cite{DN2}.

\section*{Acknowledgments}

We thank F. Essler for valuable discussions, in particular on the 
transformation (\ref{principal}) in Appendix A.
This work was supported in part by the National Science Foundation 
under Grant PHY-9870101.

\appendix

\section{Dependence on $\epsilon$}

Following many authors (see, e.g., \cite{FS},\cite{KR},\cite{TS}, ), 
we treat the full critical regime of the XXZ chain by restricting the 
anisotropy parameter $\mu$ to the range $(0 \,, {\pi\over 2})$, and 
introducing a new parameter $\epsilon = \pm 1$.  We describe here this 
approach in detail, since there are some subtleties associated with 
it, such as the dependence on $\epsilon$ in the expression 
(\ref{momentum}) for the momentum and in the boundary parameters.

\subsection{Closed chain}

We take as our starting point the following definition of the critical 
XXZ closed chain Hamiltonian
\be
{\cal H} = {1\over 4} \sum_{n=1}^{N} \left\{
\sigma^{x}_n \sigma^{x}_{n+1}
+ \sigma^{y}_n \sigma^{y}_{n+1} 
+ \cos \mu'  \left( \sigma^{z}_n \sigma^{z}_{n+1} - 1 \right) \right\} \,, 
\ee 
with $0 < \mu' < \pi$.  The repulsive regime corresponds to $0 < \mu' 
< {\pi\over 2}$, while the attractive regime corresponds to ${\pi\over 2} 
< \mu' < \pi$. The standard algebraic Bethe Ansatz procedure gives
\be
E &=& - \sin^{2} \mu' \sum_{\alpha=1}^{M} 
{1\over \cosh (2 \mu' \lambda'_{\alpha}) - \cos \mu' } \,,   \\ 
P &=& + {1\over i} \sum_{\alpha=1}^{M} 
\log {\sinh  \mu' \left( \lambda'_{\alpha} + {i\over 2} \right) 
\over \sinh  \mu' \left( \lambda'_{\alpha} - {i\over 2} \right)} 
\quad (\mbox{mod } 2 \pi) \,, 
\ee 
with
\be
\left( {\sinh  \mu' \left( \lambda'_{\alpha} + {i\over 2} \right) 
\over   \sinh  \mu' \left( \lambda'_{\alpha} - {i\over 2} \right) } 
\right)^{N} 
= \prod_{\scriptstyle{\beta=1}\atop \scriptstyle{\beta \ne \alpha}}^M 
{\sinh  \mu' \left( \lambda'_{\alpha} - \lambda'_{\beta} + i \right) 
\over   
\sinh  \mu' \left( \lambda'_{\alpha} - \lambda'_{\beta} - i \right)} 
\,, \quad \alpha = 1 \,, \cdots \,, M \,.
\ee
These are simply the formulas of Section 2 for $\epsilon=+1$ 
with primes appended to $\mu$ and $\lambda_{\alpha}$. 

The principal observation is that the ``duality'' transformation
\be
\mu &=& \pi - \mu'  \non \\
\lambda_{\alpha} &=& {\mu'\over \mu} \lambda'_{\alpha} 
+ {i \pi\over 2 \mu} 
\label{principal}
\ee 
implies
\be
{\cal H} =  {1\over 4} \sum_{n=1}^{N} \left\{
\sigma^{x}_n \sigma^{x}_{n+1}
+ \sigma^{y}_n \sigma^{y}_{n+1} 
- \cos \mu  \left( \sigma^{z}_n \sigma^{z}_{n+1} - 1 \right) \right\} \,, 
\label{app0}
\ee 
\be
E &=& + \sin^{2} \mu  \sum_{\alpha=1}^{M} 
{1\over \cosh (2 \mu \lambda_{\alpha}) - \cos \mu } \,,  \label{app1} \\ 
P &=& \pi M - {1\over i} \sum_{\alpha=1}^{M} 
\log {\sinh  \mu \left( \lambda_{\alpha} + {i\over 2} \right) 
\over \sinh  \mu \left( \lambda_{\alpha} - {i\over 2} \right)} 
\quad (\mbox{mod } 2 \pi) \,, \label{app2}
\ee 
with
\be
\left( -{\sinh  \mu \left( \lambda_{\alpha} + {i\over 2} \right) 
\over   \sinh  \mu \left( \lambda_{\alpha} - {i\over 2} \right) } 
\right)^{N} 
= \prod_{\scriptstyle{\beta=1}\atop \scriptstyle{\beta \ne \alpha}}^M 
{\sinh  \mu \left( \lambda_{\alpha} - \lambda_{\beta} + i \right) 
\over   
\sinh  \mu \left( \lambda_{\alpha} - \lambda_{\beta} - i \right)} 
\,, \quad \alpha = 1 \,, \cdots \,, M \,.
\label{app3}
\ee
The proof relies on elementary identities 
$\sinh \mu (\lambda_{\alpha} + {i\over 2}) 
= - \sinh \mu' (\lambda'_{\alpha} - {i\over 2})$,
$\sinh \mu (\lambda_{\alpha} - {i\over 2}) 
= \sinh \mu' (\lambda'_{\alpha} + {i\over 2})$, etc.

The Bethe Ansatz Eqs.  remain invariant under the transformation 
(\ref{principal}) for $N$ even.  Evidently, the attractive regime 
(${\pi\over 2} < \mu' < \pi$) corresponds to $0 < \mu < {\pi\over 2}$.  
The expressions (\ref{app1}),(\ref{app2}) coincide with the 
corresponding formulas of Section 2 with $\epsilon=-1$.  Moreover, as 
is well-known \cite{dCG}, the Hamiltonian (\ref{app0}) with $N$ even 
can be mapped by a unitary transformation to the Hamiltonian 
(\ref{hamiltonian}) with $\epsilon=-1$.

\subsection{Open chain}

We define the critical XXZ open chain Hamiltonian by
\be
{\cal H} =  {1\over 4} \left\{ \sum_{n=1}^{N-1} \left( 
\sigma^{x}_n \sigma^{x}_{n+1}
+ \sigma^{y}_n \sigma^{y}_{n+1} 
+ \cos \mu' \ \sigma^{z}_n \sigma^{z}_{n+1} \right)  
+ \sin \mu' \cot ( \mu' \xi'_{-} ) \sigma^z_1 
+ \sin \mu' \cot ( \mu' \xi'_{+} ) \sigma^z_N 
\right\} \,, \non  \\ 
\ee 
with $0 < \mu' < \pi$.  The corresponding Bethe Ansatz Eqs.  remain 
invariant under the transformation (\ref{principal}) for any $N$ 
provided there is an accompanying transformation of the boundary 
parameters,
\be
\xi_{\pm} = - {\mu' \over \mu} \xi'_{\pm} = -(\nu - 1) \xi'_{\pm} \,.
\ee 
It follows that the Hamiltonian is equal to
\be
{\cal H} =  {1\over 4} \left\{ \sum_{n=1}^{N-1} \left( 
\sigma^{x}_n \sigma^{x}_{n+1}
+ \sigma^{y}_n \sigma^{y}_{n+1} 
- \cos \mu \ \sigma^{z}_n \sigma^{z}_{n+1} \right)  
- \sin \mu \cot ( \mu \xi_{-} ) \sigma^z_1 
- \sin \mu \cot ( \mu \xi_{+} ) \sigma^z_N 
\right\} \,, \non  \\ 
\ee 
which for any $N$ can be mapped by a unitary transformation to 
\be
{\cal H} =  -{1\over 4} \left\{ \sum_{n=1}^{N-1} \left( 
\sigma^{x}_n \sigma^{x}_{n+1}
+ \sigma^{y}_n \sigma^{y}_{n+1} 
+ \cos \mu \ \sigma^{z}_n \sigma^{z}_{n+1} \right)  
+ \sin \mu \cot ( \mu \xi_{-} ) \sigma^z_1 
+ \sin \mu \cot ( \mu \xi_{+} ) \sigma^z_N 
\right\} \,. \non  \\ 
\ee 
We conclude that the critical open chain can also be described by 
the Hamiltonian (\ref{open}) with $0 < \mu < {\pi\over 2}$ and 
$\epsilon = \pm 1$, where
\be
\xi_{\pm}^{(+1)} &=& \xi'_{\pm} \non  \\
\xi_{\pm}^{(-1)} &=& -(\nu - 1) \xi'_{\pm} \,.
\ee

\section{Boundary 1-strings}

The existence of boundary string solutions of the open-chain Bethe 
Ansatz equations was discussed in Ref.  \cite{SS}.  For completeness, 
we demonstrate here the existence of boundary 1-strings following the 
approach used by Faddeev and Takhtajan \cite{FT} to study bulk 
2-strings.  We therefore consider Eq.  (\ref{BAEopen}) for the case 
$M=1$ with $N \rightarrow \infty$,

\be
\left( {\sinh  \mu \left( \lambda + {i\over 2} \right) 
\over   \sinh  \mu \left( \lambda - {i\over 2} \right) } 
\right)^{2N} 
{\sinh \mu \left( \lambda + i(\xi - {1\over 2}) \right) \over  
 \sinh \mu \left( \lambda - i(\xi - {1\over 2}) \right)} 
= 1  \,.
\ee
For simplicity, we have written boundary terms from just one boundary.
Setting $\lambda = x + i y$ with $x$, $y$ real,
\be
\left( {\sinh  \mu \left( x + i(y+{1\over 2}) \right) 
\over   \sinh  \mu \left( x + i(y-{1\over 2}) \right) } 
\right)^{2N} 
{\sinh \mu \left( x + i(y + \xi - {1\over 2}) \right) \over  
 \sinh \mu \left( x + i(y - \xi + {1\over 2}) \right)} 
= 1  \,.
\ee
Multiplying by the complex conjugate and using the identity
$\sinh (a + ib)\ \sinh (a - ib) = \sinh^{2}a + \sin^{2} b$, 
we obtain
\be
A^{2N}\ B = 1 \,,
\label{AB}
\ee
where
\be
A = {\sinh^{2}  \mu x +  \sin^{2} \mu (y + {1\over 2}) 
\over \sinh^{2}  \mu x +  \sin^{2} \mu (y - {1\over 2}) } \,, \qquad
B = {\sinh^{2}  \mu x +  \sin^{2} \mu (y + \xi - {1\over 2}) 
\over \sinh^{2}  \mu x +  \sin^{2} \mu (y - \xi + {1\over 2}) } \,.
\ee 

Evidently, there is a periodicity $y \rightarrow y + {\pi\over \mu}$.
We therefore consider two cases:
\begin{itemize}
   \item{Case I:}
\be
0 < y < {\pi\over 2 \mu} \quad (\mbox{mod } {\pi\over \mu}) 
\label{firstcase}
\ee 
It follows that 
$\sin^{2} \mu (y + {1\over 2}) > \sin^{2} \mu (y - {1\over 2})$; 
therefore $A > 1$, and hence $A^{2N} \rightarrow \infty$ for $N 
\rightarrow \infty$. Eq. (\ref{AB})  then implies $B \rightarrow 0$.
That is, for $N \rightarrow \infty$,
\be
x=0 \,, \qquad y= -(\xi - {1\over 2}) \quad (\mbox{mod } {\pi\over \mu}) 
\,. 
\ee 
The restriction (\ref{firstcase}) then implies
\be
{1\over 2}- {\pi\over 2 \mu} < \xi < {1\over 2} \quad 
(\mbox{mod } {\pi\over \mu}) \,.
\label{condition}
\ee

 \item{Case II:}
\be
-{\pi\over 2 \mu} < y < 0 \quad (\mbox{mod } {\pi\over \mu}) 
\label{secondcase}
\ee 
Then 
$\sin^{2} \mu (y + {1\over 2}) < \sin^{2} \mu (y - {1\over 2})$; 
therefore $A < 1$, and hence $A^{2N} \rightarrow 0$ for $N 
\rightarrow \infty$. Eq. (\ref{AB})  then implies $B \rightarrow \infty$.
That is, for $N \rightarrow \infty$,
\be
x=0 \,, \qquad y= \xi - {1\over 2} \quad (\mbox{mod } {\pi\over \mu}) 
\,. 
\ee 
The restriction (\ref{secondcase}) leads again to the condition 
(\ref{condition}).
\end{itemize}

In conclusion, the Bethe Ansatz Eqs. have the boundary 1-string solutions 
$\lambda = \pm i \left( \xi - {1\over 2} \right)$  
when $\xi$ satisfies the condition (\ref{condition}). 
Boundary strings of longer length are also studied in \cite{SS}.

\end{document}